\documentclass[aps,11pt,showpacs,onecolumn,nofootinbib]{revtex4}

\usepackage[utf8]{inputenc}
\usepackage{array}
\usepackage[english]{babel}
\usepackage{amsmath}
\usepackage{amsfonts}
\usepackage{amssymb}

\usepackage{hyperref}
\usepackage{subeqnarray}
\usepackage{tensor}

\usepackage{graphicx}
\usepackage{multirow}

\begin{document}

\title{Formation and Evolution of Wakes in the Spacetime Generated by a Cosmic String in $f(R)$ Theory of Gravity}
\author{G. C. Carvalho}
\email{gabriela.coutinho@fat.uerj.br}
\affiliation{Universidade Estadual do Rio de Janeiro \\ Faculdade de Tecnologia, 
27537-00, Resende - RJ, Brazil}
\author{M. E. X. Guimar\~aes}
\email{mexguimaraes@id.uff.br}
\author{P. O. Mesquita}
\email{pablomesquita@id.uff.br}
\affiliation{Instituto de F\'isica,
Universidade Federal Fluminense\\
Av. Litor\^anea S/N,
 24210-346, Niter\'oi-RJ, Brazil.}
\author{J. L. Neto}
\email{jlneto@if.ufrj.br}
\affiliation{Instituto de F\'{\i}sica, Universidade Federal do Rio de Janeiro - UFRJ \\
Av. Athos da Silveira Ramos, 149, 21941-909, Cidade Universit\'aria, Rio de Janeiro, RJ, Brazil}


\begin{abstract}
The formation and evolution of cosmic string wakes in the framework of a $f(R)$ theory of gravity are investigated in this work. We consider a simple model in which baryonic matter flows past a cosmic string. We treat this problem in the Zel'dovich approximation. We compare our results with previous results obtained in the context of General Relativity and Scalar-Theories of Gravity.
\end{abstract}

\maketitle

\section{Introduction}
Extended Theories of Gravity have recently received increasingly attention in issues such as dark matter
\cite{Rubin, Page, Capozziello, Bari} and dark energy \cite{Riess, Jailson, Simony}. In particular, the $f(R)$ theories of gravity have been suggested
as a possible alternative to explain the late time cosmic speed-up experienced
by our universe \cite{Carrol,Fay}. Such theories avoid the Ostrogradski's instability
that can otherwise prove to be problematic for general higher derivatives theories.\cite{Ostr,Woo}. Besides, these theories could be probed by the recent LIGO detections \cite{corda} and by gravitational lensing observations  \cite{liu}.

 On the other hand, it is well known that several types of topological defects may have been created by the vacuum phase transitions in the early universe \cite{kibble,vilenkin}. In particular, cosmic strings have been extensively studied in many kinds of alternative gravity theories, notably as scalar-tensor and $f(R)$ theories where many aspects and applications were developed \cite{mexg, andrade, naves, rosa}.

In this paper, our main purpose is to study the formation and evolution of wakes in cylindrically symmetric solutions in the framework of $f(R)$ theories in vacuum. In particular, we aim to explore the propagation of particles and light in a $f(R)$ cosmic string. We compare our results with the wakes formed by cosmic string solutions obtained in General Relativity and the Scalar-Tensor Theories of Gravity.

This paper is structured as follows. In Section 2, we briefly review the $f(R)$ theories of gravity (for comprehensive reviews in $f(R)$ gravity see \cite{REV}). In Section 3, for the sake of completeness, we review the obtention of the cylindrically symmetric solutions in vacuum. We give special attention to the cosmic string solution. Is it important to stress that, in this section, we follow the reference \cite{azadi}. In Section 4, the formation and evolution of wakes in the background of a $f(R)$ are studied and these are our original work. Finally, in the conclusion we summarize our main results and discuss some perspectives.

\section{$f(R)$ Gravity in the Metric Formalism - A Brief Review}

In this section, we will mainly follow the references \cite{REV}.

The action associated with the modified theories of gravity coupled with matter fields is given by:
\begin{equation}
\label{Act}
S=\frac{1}{2\kappa}\int d^{4}x\sqrt{-g}f(R)+{\cal S}_m \ ,
\end{equation}
where $f(R)$ is an analytical function of the the Ricci scalar, $R$, $\kappa=8\pi G$ and ${\cal S}_m$ corresponds to the action associated with the matter fields. By using the metric formalism, the field equations become:
\begin{equation}
\label{G}
G_{\mu\nu}\equiv R_{\mu\nu}-\frac{1}{2}Rg_{\mu\nu}=T^{c}_{\mu\nu}+\kappa \tilde{T}^{m}_{\mu\nu}	\ ,
\end{equation}
in which $T^{c}_{\mu\nu}$ is the geometric energy-momentum tensor, namely
\begin{eqnarray}
\label{3}
&&T^{c}_{\mu\nu}=\frac{1}{F(R)}\left\{\frac{1}{2}g_{\mu\nu}\left(f(R)-F(R)R\right)\right. \nonumber\\
&+&\left. \nabla^{\alpha} \nabla^{\beta} F(R)\left(g_{\alpha\mu}g_{\beta\nu}-g_{\mu\nu}g_{\alpha\beta}\right)\right\}
\end{eqnarray}
with $F(R)\equiv\frac{df(R)}{dR}$.

The standard minimally coupled energy-momentum tensor, ${T}^{m}_{\mu\nu}$, derived from the matter action, is related to $\tilde{T}^m_{\mu\nu}$ by
\begin{equation}
\label{T}
\tilde{T}^{m}_{\mu\nu}=T^m_{\mu\nu}/F(R) \ .
\end{equation}
Thus the field equations can be written as
\begin{eqnarray}
\label{FE}
&&F(R)R_{\mu\nu}-\frac{1}{2}f(R)g_{\mu\nu}-\nabla_{\mu}\nabla_{\nu}F(R)+g_{\mu\nu}\Box F(R)=\kappa T^{m}_{\mu\nu}\ .
\end{eqnarray}
Taking the trace of the above equation we get
\begin{equation}
\label{6}
F(R)R-2f(R)+3\Box{F(R)}=\kappa T^{m} \ ,
\end{equation}
which express a further scalar degree of freedom that arises in the modified theory. Through this equation it is possible to express $f(R)$ in terms of its derivatives and the trace of the matter energy-momentum tensor, as follows
\begin{eqnarray}
\label{fr}
f(R)=\frac{1}{2}\left(F(R)R+3\Box{F(R)}-\kappa T^{m}\right) \ .
\end{eqnarray}
Substituting the above expression into (\ref{FE}) we obtain
\begin{equation}
F(R)R_{\mu\nu}-\nabla_{\mu}\nabla_{\nu}F(R)-\kappa \tilde{T}^{m}_{\mu\nu}= \frac{g_{\mu\nu}}{4}\left[F(R)R-\Box F(R)-\kappa T^{m}\right] \; .
\end{equation}
From this expression we can see that the combination below
\begin{equation}
\label{A0}
	C_\mu=\frac{F(R)R_{\mu\mu}-\nabla_{\mu}\nabla_{\mu}F(R)-\kappa T^{m}_{\mu\mu}}{g_{\mu\mu}} \ ,
\end{equation}
with fixed indices, is independent of the corresponding index. So, the following relation
\begin{equation}
\label{Ceq}
C_{\mu}-C_{\nu}=0\ ,
\end{equation}
holds for all $\mu$ and $\nu$.

\section{Vacuum Cylindrically Symmetric Solutions in $f(R)$ Gravity: A Brief Review}
\label{sec3}
In this section we review how to derive the field equations associated with cosmic string system in the context of modified theories of gravity following Azadi et al. \cite{azadi}.

The $f(R)$ field equations in vacuum are given by

\begin{equation}
\label{FEV}
F(R)R_{\mu\nu}-\frac{1}{2}f(R)g_{\mu\nu}-(\nabla_{\mu}\nabla_{\nu}-g_{\mu\nu}\Box ) F(R)=0
\end{equation}

Taking the trace of eq. (\ref{FEV}), we have

\begin{equation}
\label{traceV}
F(R)R- 2f(R)+3\Box F(R)=0 \; .
\end{equation}

Now, since we are interested in obtaining static solutions with cylindrical symmetry in vacuum, we will work with a general metric in Weyl coordinates $(t,r,\phi, z)$ given by

\begin{equation}
\label{weyl}
g_{\mu\nu}=diag(-e^{2k-2u}, e^{2k-2u},\omega^2e^{-2u}, e^{2u}) \;\; ,
\end{equation}
where $k, u$ and $\omega$ are functions of $r$ only.

The non-zero components of the Ricci tensor are:

\begin{eqnarray}
\label{ricci}
R_{00} & = & k'' - u'' + \frac{k'\omega'}{\omega} - \frac{u'\omega'}{\omega} \nonumber \\
R_{11} & = & -k'' + u'' - \frac{\omega ''}{\omega} + \frac{k'\omega'}{\omega} - \frac{u'\omega'}{\omega} - 2u'^2 \nonumber \\
R_{22} & = & e^{-2k}(\omega\omega 'u'- \omega\omega '' + \omega^2 u'') \\
R_{33} & = & e^{4u-2k} (-u'' - \frac{u'\omega'}{\omega}) \nonumber \;\; ,
\end{eqnarray}
where prime $(')$ indicates derivative with respect to $r$. Therefore, the scalar curvature is

\begin{equation}
\label{scalarricci}
R = -2e^{2u} \left( \frac{-\omega u'' +\omega k'' - u'\omega '+ \omega '' +\omega u'^2}{\omega e^{2k}} \right)
\end{equation}

Replacing $f(R)$ of eq. (\ref{traceV}) in eq. (\ref{FEV}), finally we get:

\begin{equation}
\frac{FR_{\mu\nu} - \nabla_{\mu}\nabla_{\nu}F}{g_{\mu\nu}} = \frac{1}{4} (FR - \Box F(R)) \; .
\end{equation}
Defining $A_{\mu}$ (which is the equivalent of $C_{\mu}$ in the absence of matter) as

\begin{equation}
\label{amu}
A_{\mu} = \frac{FR_{\mu\mu} - \nabla_{\mu}\nabla_{\mu}F}{g_{\mu\mu}} \; ,
\end{equation}
we can easily see that

\begin{equation}
\label{amu2}
A_{\mu} = \frac{1}{4} (FR - \Box F(R)) \; ,
\end{equation}
which is a scalar quantity. This means that $A_{\mu}=A_{\nu}$ for any $\mu, \nu$, which implies that we can replace
 eq. (\ref{FEV}) by $A_t= A_r$, $A_t=A_{\phi}$ and $A_t= A_z$. After some straightforward calculation, we finally get, respectively

\begin{eqnarray}
\label{amu3}
- F'' + 2 F'(k'- u') +F \left( - \frac{2k'\omega '}{\omega} + \frac{\omega ''}{\omega} + 2u'^2  \right) & = & 0 \\
F\omega^2  ( -k'' - \frac{k'\omega '}{\omega}+ \frac{\omega ''}{\omega} ) + F'(\omega\omega '- \omega^2 k') & = & 0 \\
F \left(-k''+ 2u'' - \frac{k'\omega '}{\omega}  + \frac{2\omega 'u'}{\omega} \right) + F'(k'-2u') & = & 0.
\end{eqnarray}
Therefore, any group of functions $k,u,$ and $\omega$ which satisfies the equations above is a solution of the modified gravity equations in vacuum.

Since these equations are highly nonlinear, we will consider the particular case where $R=0$. This is justified by the fact that we are interested in cosmic string solution and the external metric of a cosmic string is locally flat.

\subsection{Field Equations Solutions for the Special Case $R=0$}

Let us start deriving eq. (\ref {FEV}) with respect to $r$,

\begin{equation}
\label{FEVD}
RF'- R'F + 3 (\Box F)'=0 \; .
\end{equation}

In the particular case where $R=$ constant, eq. (\ref{FEVD}) implies that $F'=0$. As a consequence, eqs. (17-19) become

\begin{eqnarray}
\label{zeroR}
2u'^2 + \frac{\omega ''}{\omega} - 2\frac{k'\omega '}{\omega} & = & 0 \\
k'' + \frac{k'\omega '}{\omega} -\frac{\omega ''}{\omega} & = & 0 \\
2u'' + 2 \frac{\omega 'u'}{\omega} - \frac{k'\omega '}{\omega} - k'' & = & 0
\end{eqnarray}

In order to solve this non-linear equations system, let us first sum up eqs. (22) and (23). In doing this, we get

\begin{equation}
\label{sum}
u'' + \frac{\omega 'u'}{\omega} - \frac{1}{2}\frac{\omega ''}{\omega} = 0 \; .
\end{equation}
Defining $u'= g(r)$ in eq. (\ref{sum}), we get

\begin{equation}
\label{new}
\omega g'+ \omega 'g = \frac{1}{2} \omega ''
\end{equation}
which can be rewritten as
\begin{equation}
\label{sum1}
\frac{d(\omega g)}{dr} = \frac{1}{2} \frac{d}{dr}\left( \frac{d\omega}{dr} \right) \; .
\end{equation}
Integrating eq. (\ref{sum1}), we obtain

\begin{equation}
\label{sum2}
g(r) = u'= \frac{1}{2} \frac{\omega '+ c_2}{\omega} \; ,
\end{equation}
where $c_2$ is a constant to be determined later.

Now, subtracting eq. (23) from eq. (22), we get
\begin{equation}
\label{sub}
k'' + \frac{k'\omega '}{\omega} - \frac{\omega ''}{\omega} = 0 \; ,
\end{equation}
and we find
\begin{equation}
\label{sub1}
k'= \frac{\omega ' + c_1}{\omega}
\end{equation}
where $c_1$ is a constant to be determined later.

Replacing eqs. (\ref{sum2}) and (\ref{sub1}) in eq. (21), we get a differential equation for the function $\omega$, which is

\begin{equation}
\label{omega}
\frac{1}{2} \left(\frac{\omega '+ c_2}{\omega}\right)^2 + \frac{\omega ''}{\omega} = 2\omega '\frac{(\omega '+ c_1)}{\omega ^2}
\end{equation}

In order to solve our equations, we will make the hypothesis that $\omega(r)$ is a linear function of $r$. This is justified by cosmic string solutions in either General Relativity or Scalar-tensor gravities. Therefore, $\omega ''=0$. Hence, we have

\begin{eqnarray}
u & = & c_3 \pm \sqrt{\frac{c_5}{c_6}} \ln \omega \\
k & = & c_4 + \frac{c_5}{c_6} \ln {\frac{\omega}{c_6}} \\
\omega & = & c_6 r + c_7
\end{eqnarray}

In order to satisfy eq. (\ref{omega}), the constants $c_5$ and $c_6$ must obey the following relations

\begin{eqnarray}
\label{constantes}
c_2 = (2 \sqrt{\frac{c_5}{c_6}} - 1)c_6 \nonumber \\
c_1 = c_5 - c_6 \; .
\end{eqnarray}
We can easily see that the metric functions (33-35) satisfy eq. (\ref{omega}) in the particular case where $R=0$.

Redefining the quantities $\tilde{c_4} = c_4 - (c_5/c_6)\ln c_6$ and $\rho = \omega = c_6 r$ and making $c_7=0$ without any loss of generality, we can write down the metric in Weyl coordinates as

\begin{equation}
\label{metric1}
ds^2 = e^{-2(c_3 \pm \sqrt{\frac{c_5}{c_6}}\ln \rho)}[e^{2(\tilde{c_4}+\frac{c_5}{c_6}\ln\rho)}\left(\frac{d\rho^2}{c_6^2} - dt^2\right) + \rho^2 d\phi^2] + e^{2(c_3 \pm \sqrt{\frac{c_5}{c_6}}\ln \rho)}dz^2 \; .
\end{equation}
Defining the quantities $m = \sqrt{\frac{c_5}{c_6}}$ and $A= \frac{e^{\tilde{c_4}- c_3}}{c_6}$ and defining new coordinates such as

\begin{eqnarray}
\tilde{t} & = & c_6 A^{\frac{1}{m(m\mp 1) +1}} t \nonumber \\
\tilde{\rho} & = & A^{\frac{1}{m(m\mp 1) +1}} \rho \nonumber \\
\tilde{\phi} & = & e^{-c_3}A^{-\frac{1 \mp m}{m(m\mp 1) +1}}\phi \\
\tilde{z} & = & e^{c_3}A^{\frac{\mp m}{m(m\mp 1) +1}}z \; .
\end{eqnarray}

In doing this, the metric reduces to a very simple form\footnote{The calculations are long but straightforward.}

\begin{equation}
\label{metric2}
ds^2 = \tilde{\rho}^{2m(m\mp 1)}(-d\tilde{t}^2 + d\tilde{\rho}^2) + \tilde{\rho}^{2(1 \mp m)}d\tilde{\phi}^2 + \tilde{\rho}^{\pm 2m} d\tilde{z}^2
\end{equation}

Applying the complex transformation $\tilde{t} \rightarrow i\tilde{z}$ and $\tilde{z} \rightarrow i\tilde{t}$, we get a well known metric \cite{Hiscock}

\begin{equation}
\label{metric3}
ds^2 = \tilde{\rho}^{2m(m\mp 1)}(d\tilde{z}^2 + d\tilde{\rho}^2) + \tilde{\rho}^{2(1 \mp m)}d\tilde{\phi}^2 - \tilde{\rho}^{\pm 2m} d\tilde{t}^2 \; ,
\end{equation}
which, apart from the sigh $\mp$ is pretty much the same as the Levi-Civita static cylindrically symmetric solution in General Relativity \cite{LC}.

When $m=0$, the spacetime (\ref{metric3}) becomes\footnote{We will suppress the $\tilde{}$
symbol from now on because it will not cause any confusion.}

\begin{equation}
\label{metric4}
ds^2 = -dt^2 + d\rho^2 + \rho^2e^{-2c_4}c_6^2d\phi^2 + dz^2 \; .
\end{equation}
It is very easy to see that this spacetime is locally flat but not globally Euclidean. This spacetime  is conical with a deficit angle equal to

\begin{equation}
\label{deficit}
\delta\phi = 2\pi (1 - e^{-2c_4}c_6) \; ,
\end{equation}
as long as $e^{-2c_4}c_6 <1$ which imposes some constraints on the constants $c_4$ and $c_6$.

\begin{figure}[h!]
\begin{center}
\includegraphics[width=10cm]{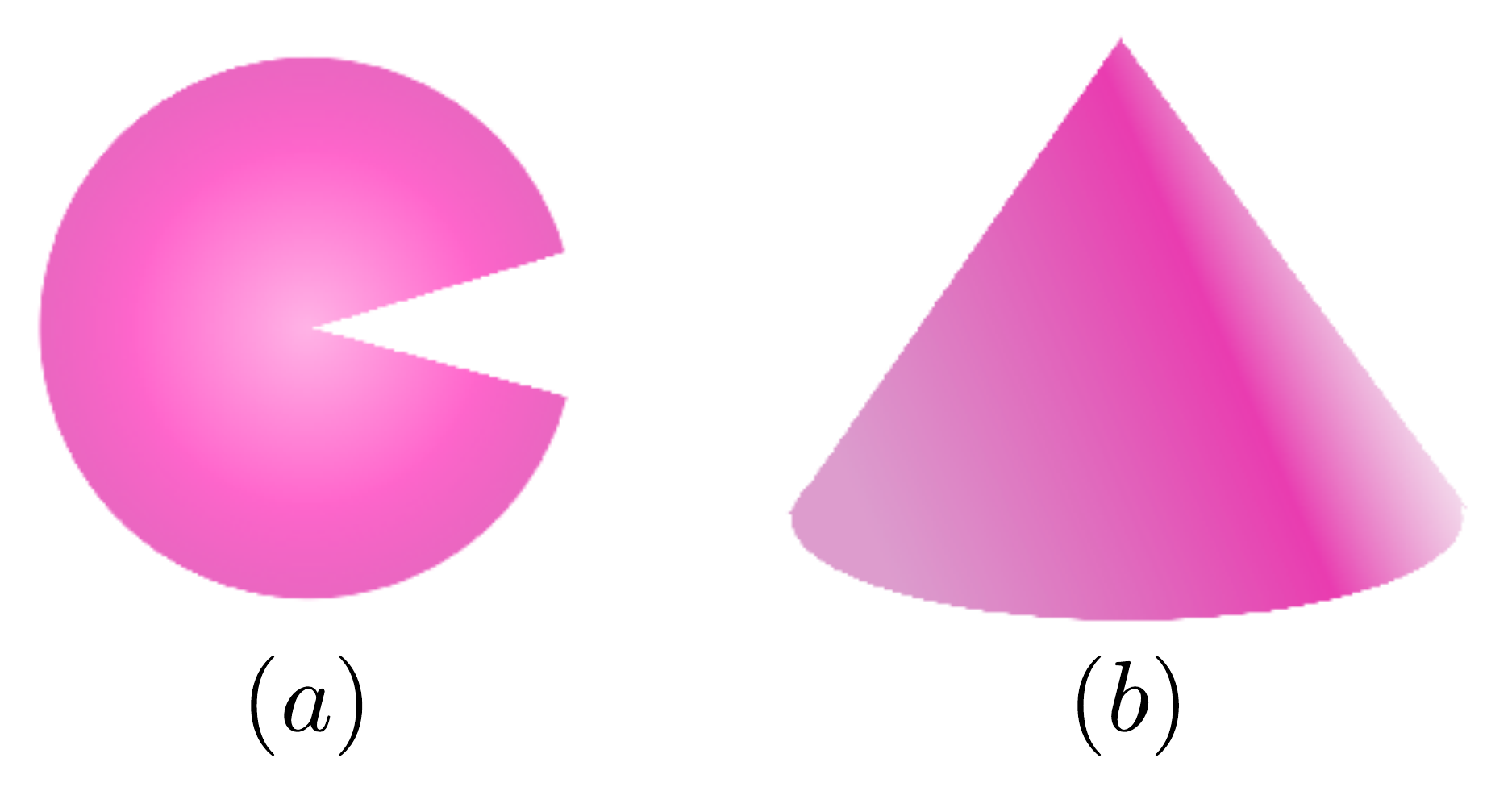}
\caption{The Conical Spacetime.}
\label{figure1}
\end{center}
\end{figure}

\section{Formation and Evolution of Wakes in The Spacetime of a $f(R)$ Cosmic String}

In this section we will study the formation and evolution of some structures when a cosmic string moves through a region containing baryonic matter.

\subsection{The Formation of Wakes}
Let us suppose that the cosmic string moves with a constant velocity $v_c$ in the  $x$-axis. Since $g_{tt}$ is constant in (\ref{metric4}), the string does not exert any gravitational force on test particles. However, test particles do suffer a perturbation when passing through a cosmic string. To see that, let us consider that one is in a comoving frame in which the string is at rest and, by a Lorentz transformation, the baryonic matter is approaching the string with the velocity $v_c$. From Fig. 2, we can see how the velocity of the test particles are perturbed

\begin{figure}[h!]
\begin{center}
\includegraphics[width=10cm]{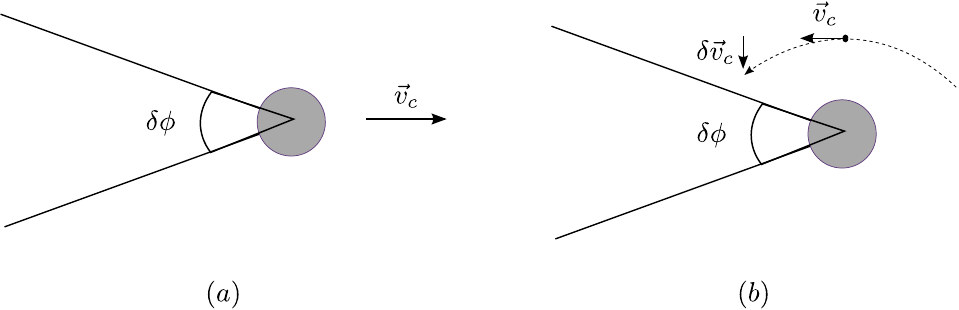}
\caption{(a) A Cosmic String Moving with Constant Velocity $v_c$. (b) Particles Moving with Constant Velocity $v_c$ in the Comoving String Frame.}
\label{perturbation}
\end{center}
\end{figure}
and  we can calculate this perturbation. Of course, it depends on the deficit angle and, hence, on the parameters of the $f(R)$ theory
\begin{equation}
\label{ui}
u_i = \delta v_c \approx \frac{\delta \phi}{2} \gamma v_c = e^{-2c_4}c_6 \pi \gamma v_c \; ,
\end{equation}
where $\gamma = (1-v_c^2)^{-1/2}$, considering $c=1$.

In this way, particles which move in regions where $y>0$ may collide with particles which move in regions where $y<0$ after passing through the string and form stable structures called "wakes", see Fig. 3.

\begin{figure}[h!]
\begin{center}
\includegraphics[width=15cm]{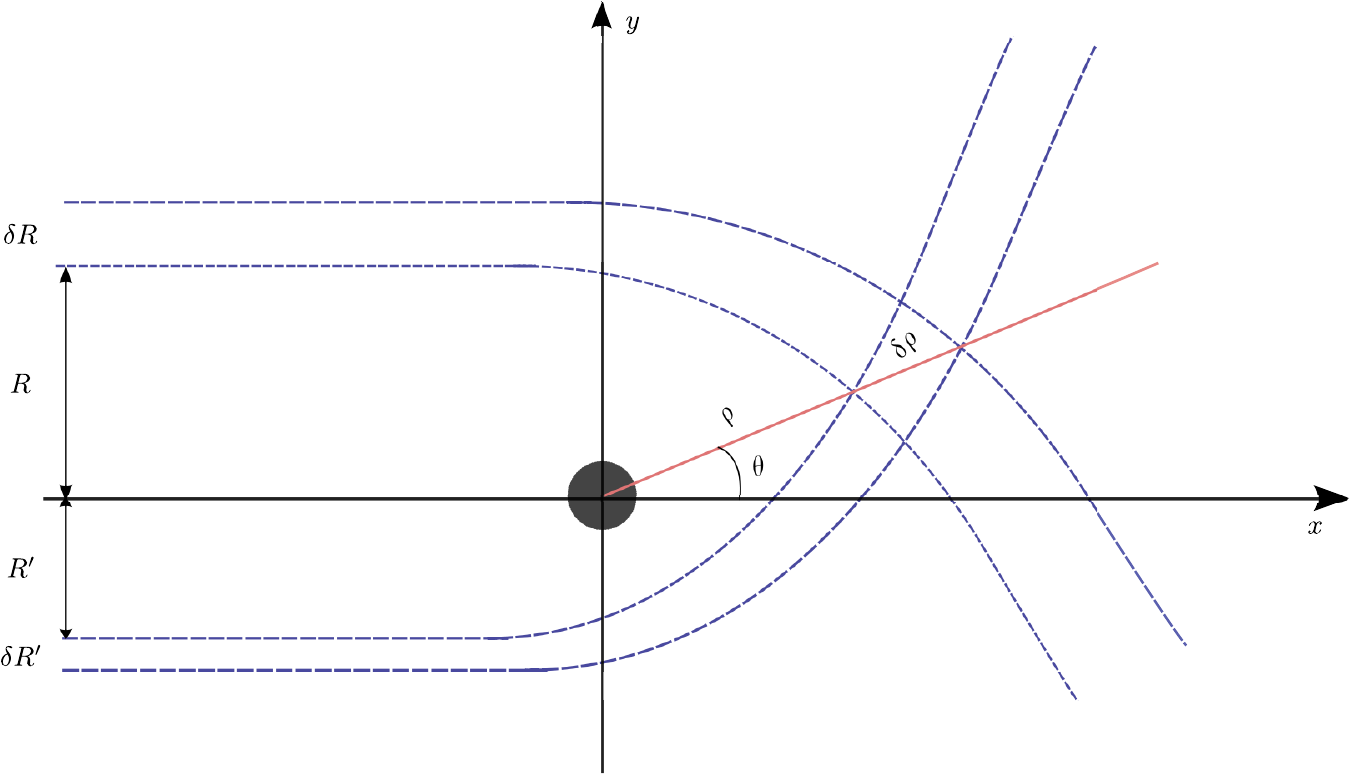}
\caption{Particles Collision with Impact Parameters equal to $R$ and $R'$.}
\label{impacto}
\end{center}
\end{figure}

\subsection{The Evolution of Wakes: The Zel'dovich Approximation}

Let us now make a quantitative description of the accretion problem using the Zel'dovich approximation \cite{aguirre}, which consists in considering the Newtonian accretion problem in an expanding Universe by means of the method of linear approximation.

Suppose that the wake is formed at $t_i > t_{eq}$, where $t_{eq}$ is the time in which matter starts to dominate over radiation. The physical trajectory of a particle can be written as

\begin{equation}
\label{trajectory}
{\vec{r}}({\vec{q}},t) = a(t) ({\vec{q}} - {\vec{\psi}}({\vec{q}},t)) \; ,
\end{equation}
where ${\vec{q}}$ is the unperturbed comoving position and ${\vec{\psi}}$ is the comoving displacement developed as a consequence of the gravitational attraction induced by the wake on the particle.

The equation of motion in the Newtonian limit is given by

\begin{equation}
\label{eqnewton}
\ddot{\vec{r}}= -\frac{\partial\Phi ({\vec{r}}, t)}{\partial{\vec{r}}} \; ,
\end{equation}
where $\Phi({\vec{r}}, t)$ is the Newtonian potential which obeys the Poisson equation\footnote{In order to avoid confusion, we have changed the radial coordinate for $r$ instead of $\rho$ which represents now the matter density.}

\begin{equation}
\label{poisson}
\nabla^2\Phi = 4 \pi  G \rho ({\vec{r}},t) \; .
\end{equation}

The matter density $\rho ({\vec{r}},t)$ is determined in terms of the background density $ \rho_0(t)$ as \cite{aguirre}

\begin{equation}
\label{densidade}
\rho ({\vec{r}},t) d^3{\vec{r}} = a^3 (t) \rho_0(t) d^3{\vec{q}}
\end{equation}
which gives

\begin{equation}
\label{densi}
\rho ({\vec{r}},t) \simeq \rho_0 \left(1 + \frac{\partial\psi_r (q,t)}{\partial q}\right) \; ,
\end{equation}
where $r, q$ and $\psi_r$ are the radial components of these quantities.

Replacing eq. (\ref{densi}) in the Poisson equation (\ref{poisson}), we obtain

\begin{equation}
\label{poissonlin}
\frac{\partial\Phi}{\partial{\vec{r}}}=4\pi G [\frac{\rho_0(t)}{3}\vec{r} +\rho_0(t)a(t){\vec{\psi}}(\vec{q} ,t)] \; .
\end{equation}

 If we replace the relation $\ddot{a}/a = -4\pi G \rho_0/3$ in the equation above, we get the linearized equation for $\psi$

\begin{equation}
\label{psi}
\ddot{\psi} + 2\frac{\dot{a}}{a}\dot{\psi} +3\frac{\ddot{a}}{a}\psi =0 \; .
\end{equation}

Since we are working in the matter era, $a(t) \propto t^{2/3}$. Hence, eq. (\ref{psi}) becomes

\begin{equation}
\label{psi2}
\ddot{\psi} + \frac{4}{3t}{\dot{\psi}} - \frac{2}{3t^2}\psi =0 \; .
\end{equation}
The equation above is the well-known Euler equation. Applying appropriate conditions such as $\psi(t_i)=0$ and $\dot{\psi(t_i)}=-u_i$ the solution is

\begin{equation}
\label{euler}
\psi (x,t) = \frac{3}{2} [\frac{u_i t_i^2}{t} - u_i t_i(t/t_i)^{2/3}] \; .
\end{equation}

The comoving coordinate $q(t)$ can be calculated using the fact that $\dot{r} =0$, which means that, eventually, the particle stops expanding with the Hubble flow and starts to collapse onto the wake. Therefore, we get

\begin{equation}
\label{q}
q(t)= -\frac{6}{5} [\frac{u_i t_i^2}{t} - u_i t_i(t/t_i)^{2/3}] \; .
\end{equation}

Hence, we are now able to compute the wake's thickness $d(t)$ and surface density $\sigma(t)$ \cite{vacha}

\begin{eqnarray}
d(t) & \approx & 2 q(t) \left(\frac{t}{t_i}\right)^{2/3} \\
\sigma(t) & \approx & \rho_0(t) d(t) \; .
\end{eqnarray}

Finally, we obtain

\begin{eqnarray}
d(t) & \approx & \frac{12}{5} a_0 \pi \gamma v_c \left[\frac{t^{4/3}}{t_i^{1/3}} - \frac{t_i^{4/3}}{t^{1/3}}\right]^{2/3} \\
\sigma & \approx & \frac{12}{5}\rho_0 a_0 \pi \gamma v_c \left[\frac{t^{4/3}}{t_i^{1/3}} - \frac{t_i^{4/3}}{t^{1/3}}\right]^{2/3}
\end{eqnarray}
where $a_0 = e^{-c_4}c_6$, which means that the wake's thickness and density depend on the parameters of the $f(R)$ theory.

\section{Final Remarks}

In this work we considered cylindrically symmetric solutions in the framework of $f(R)$ theory of gravity. These solution were obtained in vacuum regime and in the particular case where $R=0$. A cosmic string solution was of special interested and we studied the formation and evolution of wakes in this spacetime. Comparing our results with those obtained previously in the literature \cite{vacha, masalskiene}, both in General Relativity and in Scalar-Tensor theories, respectively, we can see that they ressemble with the GR wakes instead of the scalar-tensor ones, as we would expected since $g_{tt}$ is constant and there is no gravitational force exerted by the $f(R)$ cosmic string in the same way as the GR cosmic string.  However, for a precise  comparison and further numerical evaluation, we must consider the internal cosmic string matter configuration because we need to determine the metric constants $c_4 , c_6$. In particular, they must obey the GUT cosmic string order of magnitude for all parameters. 

As we expected, all wake's physical quantities depended on the parameters of the particular theory of gravity under consideration. But, again, in order to make a quantitative evaluation, we need to consider not a vacuum solution but  a full energy-momentum tensor for the internal cosmic string configuration in the same way as \cite{mexg, garfinkle}. This is under consideration now.

As a future perspective of this work, we plan to study the $f(R)$ cosmic string as a generator of the rotational curves in galaxies \cite{Rubin}. This work will come as a forthcoming paper.

\section*{Acknowledgments}
P. O. Mesquita and M. E. X. Guimar\~aes would like to thank PIBIC/CNPq (Conselho Nacional de Desenvolvimento Cient\'{\i}fico e Tecnol\'ogico) for a support during the preparation of this work.

\end{document}